Sep. 30

# Selection of scheduling Algorithm

Prof S. Yashvir, Om Prakash
Department of Computers & Statistics

## Abstract

The objective of this paper is to take some aspects of disk scheduling and scheduling algorithms. The disk scheduling is discussed with a sneak peak in general and "selection of algorithms" in particular.

## 1. Introduction

The Disk is said to be of two basic types:

1. Fixed head disk- This has one head for each track on the disk and it requires no head movement time to service a request. This is quite expensive.
2. Movable head disk- This is much more common in use because it has a single head driven by a stepper motor that can position the head over any desired track on the disk surface.

The task of scheduling usage of sharable resources by the various processes is one of the important jobs of operating system as it is responsible for efficient use of the disk drives.

The efficiency of disk drivers means that disks must have fast access time and reasonable bandwidth. The two major components of access time and bandwidth of disks are:

- ■ *Seek time*-the time to move the heads to the cylinder containing the desired sector.
- ■ *Rotational latency*-the additional time to rotate the desired sector to the disk head.

This can be considered very important in case of systems with multi programming as they have a common file system. The file system is said to be common in multi programmed systems because it is shared by all the users even though each of them may have one's own file. This common file system may be spread out over a finite number of disks or it may reside entirely on a single disk.

Thus, all processes that do disk IO are competing for access to the same physical disk or set of physical disks. Mostly as any given disk can only perform one access at a particular time, if several accesses are requested on a given disk, some order of service for the requests is established by the OS.

The only exception is that there are two or more independent head assemblies on some disks and so those can perform two or more service requests at a single time. However, even in this type of cases too, scheduling is a must if there are more requests outstanding than the available heads to serve them.

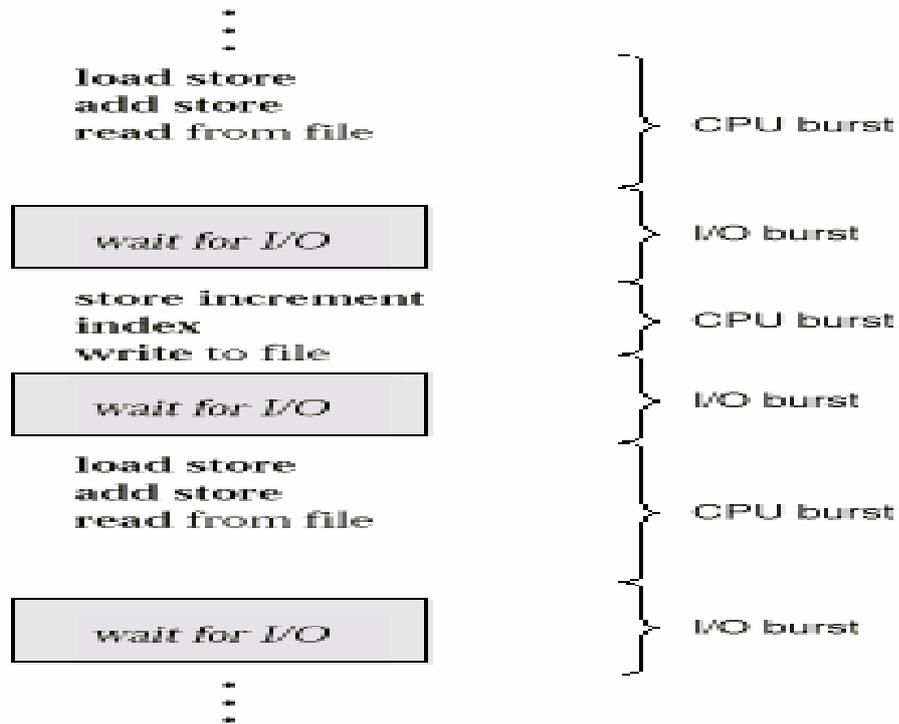

## 2. Disk Scheduling

It is now clear that there can be number of programs in memory at the same time that results in overlapping of CPU and I/O.

There are batch programs that run without interaction from user. There may be time shared programs that run with user interaction. For both of these the common name used is Process for which burst cycle of CPU characterizes execution of their process, alternatively between CPU and I/O activity. The scheduling makes selection among the processes in memory that are ready to be executed and makes allocation of the CPU to one of them. The decision regarding scheduling takes place when a process switches from:

- running to waiting state
- running to ready state
- waiting to ready state
- Terminates

The scheduling of the above processes is known as nonpreemptive. It must be noted that mostly the scheduling quantum is not used by almost all processes as shown below:

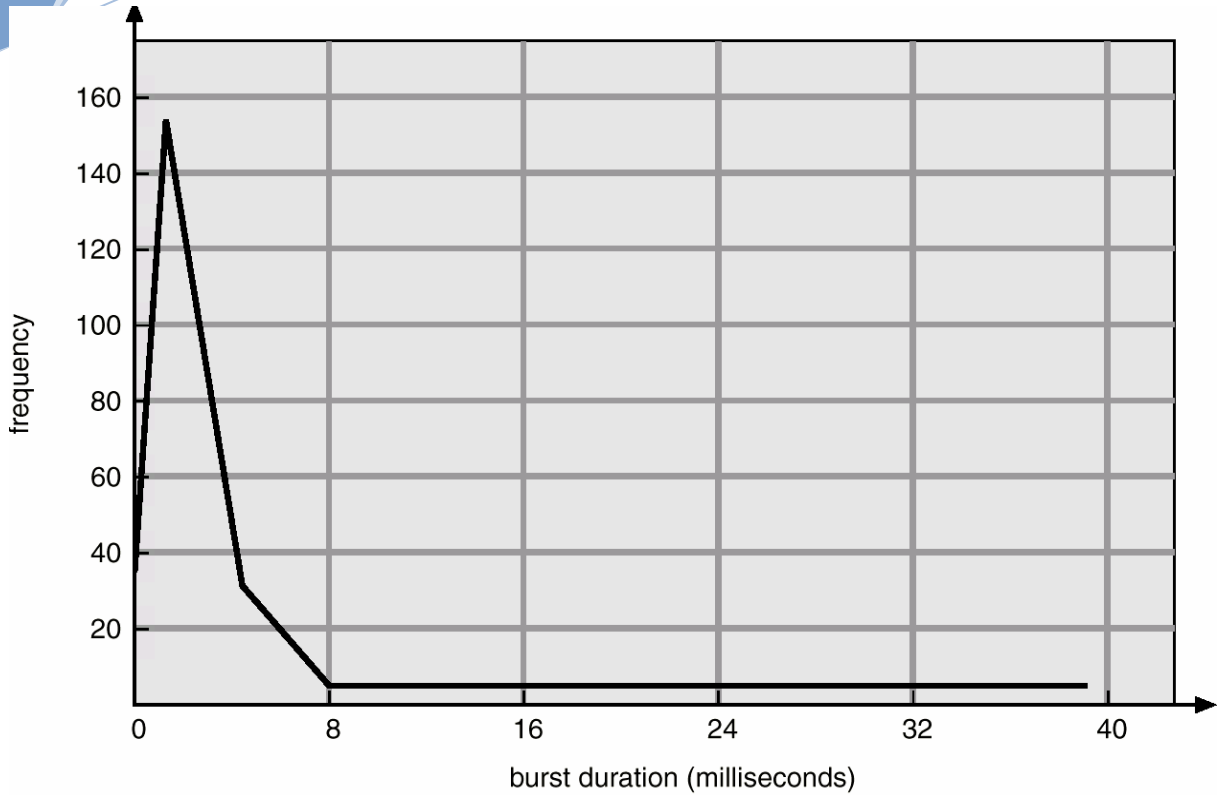

## 3. Scheduling Algorithms

### 3.1    First Come First Serve (FCFS)

It is similar to FIFO. It is simple, fair approach but perhaps not the best because of its poor performance as average queue time may be too long to be served. It is quite difficult to find the average queue and residence times for this. Of course, the simplest .

way but if disk accesses are scheduled in an order that takes into consideration some of the physical characteristics of the disk then system can be improved significantly throughout. For example, for the following processes request queue 98, 183, 37, 122, 14, 124, 65, 67, with head pointer 53, total head movement is 640 cylinders

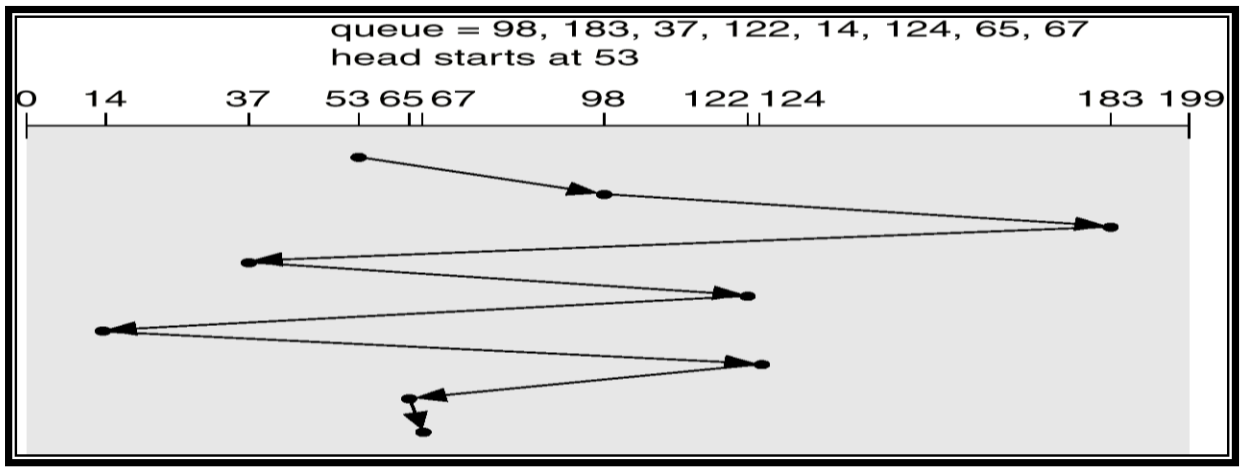

### 3.2    SSTF: Shortest Seek Time First

It is much more efficient, but leads to starvation. It may be optimal for minimizing queue time, but may be impossible to be implemented as it tries to predict the scheduled process based on previous history.  It selects the request with the minimum seek time from the current head position. It is a form of SJF scheduling; may cause starvation of some requests.

The prediction of the time used by the process on its next schedule can be given by

$t( n+1 ) = w * t( n ) + ( 1 - w ) * T( n )$

Where, $t(n+1)$ is time of next burst.
$t(n)$ is time of current burst.
$T(n)$ is average of all previous bursts
$W$ is a weighting factor emphasizing current or previous bursts.

For Example, with head pointer at 53, Total head movement:

$98 + 183 + 37 + 122 + 14 + 124 + 65 + 67 = 236$ tracks or $<$ 30 tracks per access

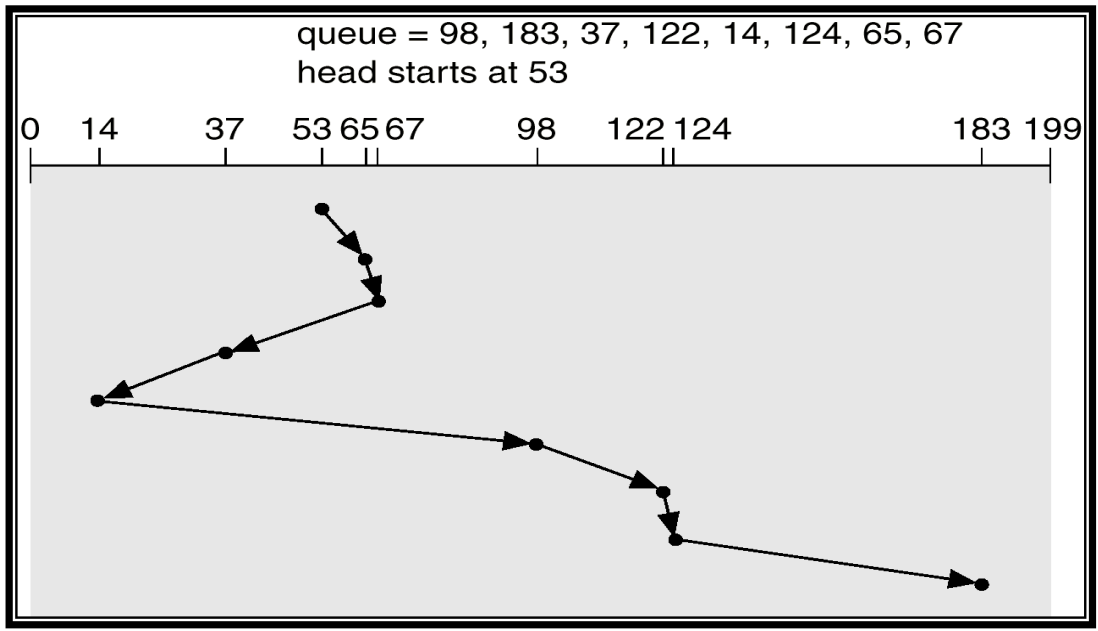

But SSTF can be a problem on a heavily used disk. If one request is at the extreme and the other request is nearer to the centre, the extreme request can be postponed for a long time.

### 3.3    SCAN

The purpose of it is to combine efficiency with fairness. The process starts at one end of the disk with movement towards the other end, servicing requests until end, where the head movement is reversed and servicing continues. It is also known as the elevator algorithm because of its working similar to an elevator services in a building. When it goes up, it requests services in order from floors above it, but floors below it, are ignored. When it goes down, it only requests services below it. For

Sep. 30

example, with head pointer 53, Total head movement is:

98 +183 + 37 + 122 + 14 + 124 + 65 + 67 = 208 tracks

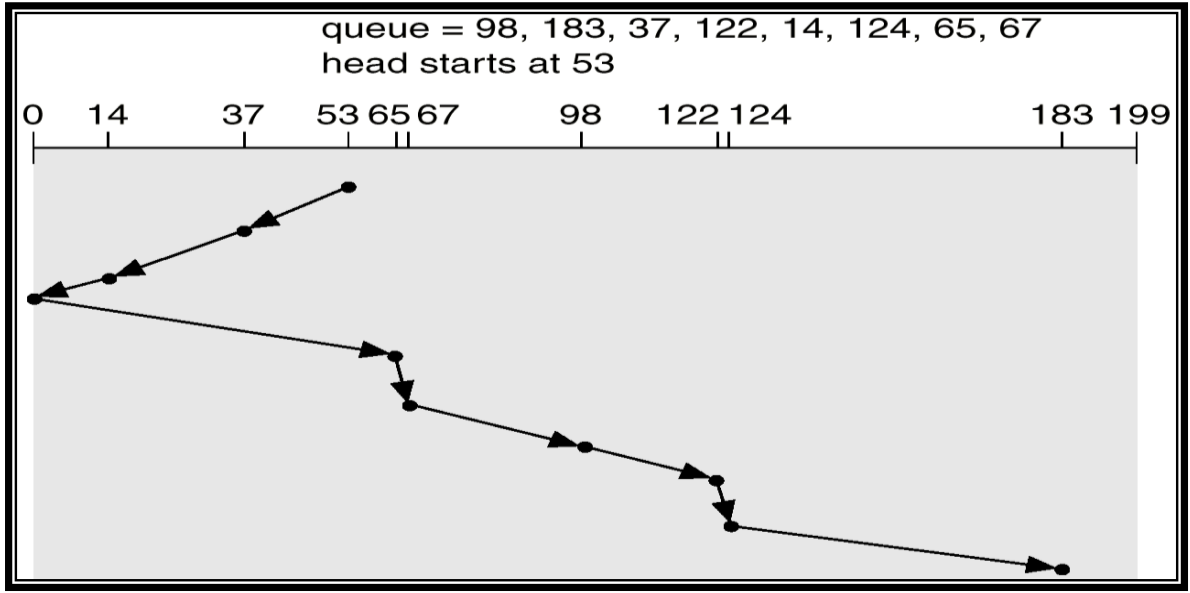

## 3.4 C-SCAN: Circular SCAN

This algorithm is similar to SCAN. The only exception is that the disk requests services in one direction only and "jumps" to the starting of disk when the last track is reached. This results in a more uniform response time. Since a single large jump may be faster than several smaller ones, overall it may be more efficient than SCAN. By providing a more uniform wait time and treating the cylinders as a circular list, it proves better than SCAN. For example, with head pointer 53,

Total head movement:

98 +183 + 37 + 122 + 14 + 124 + 65 + 67 = 322 tracks

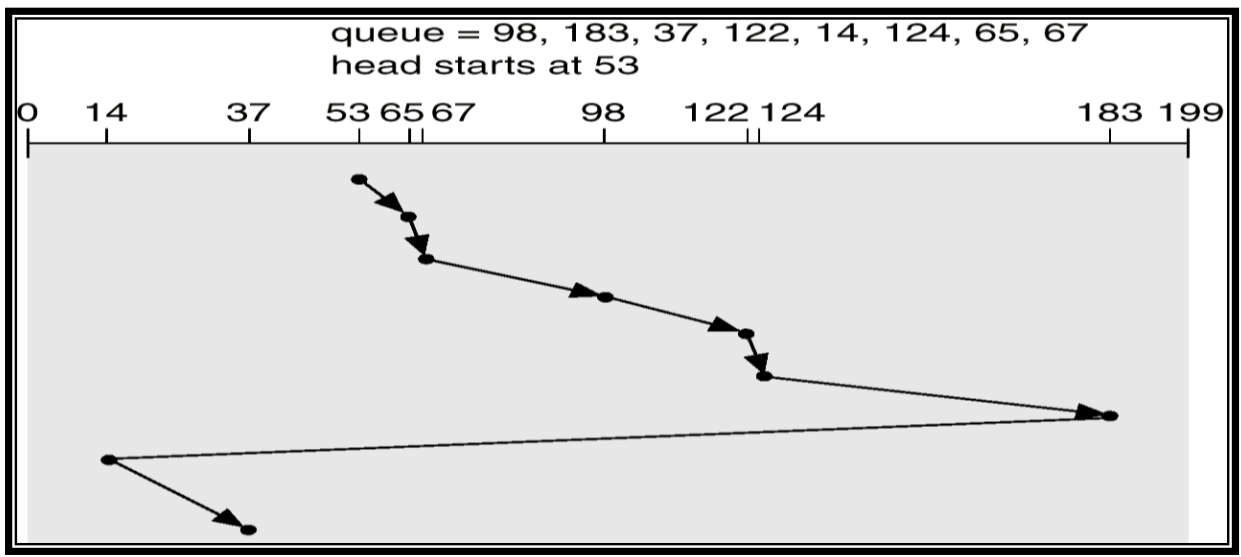

Sep. 30

## 3.5 FSCAN

With the above discussed algorithms it may be possible that the arm may not move for a considerable period of time. To avoid this arm stickiness the disk request queue can be segmented, with one segment being processed at a time completely.

FSCAN is an example of such an approach. It is the policy that uses two sub queues. When a SCAN begins, all of the requests are in one of the queues, with the other empty. During the scan, requests are put into the performance approaches similar to SCAN and for N=1, it approaches FIFO.

## 3.7 **MULTI-LEVEL QUEUE:**

This type of algorithm has multi queues with each queue having its own algorithm. Then priority based other queue. This means that till all the old requests gets processed, service of the new requests is deferred.

## 3.6    N-step-SCAN

This policy segments the queue of disk request into sub queues of length N and the processing of these is one at a time using SCAN. Till the processing of a queue, new requests are added to some other queue. If requests available are less than N, at the end of scan, then all of them are processed with the next scan. With larger values of    N,                          the algorithm arbitrates between those multi level queues that can use feedback to move between queues. This method is flexible but complex. For example

highest priority

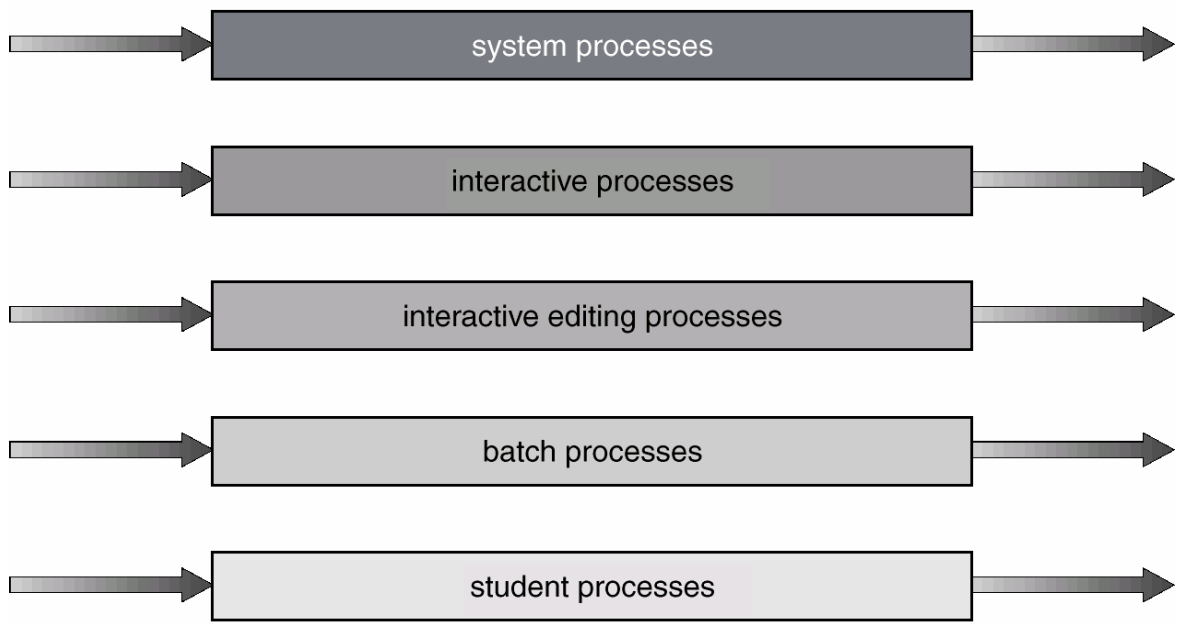

lowest priority

## 3.8 MULTIPLE PROCESSOR SCHEDULING:

We know that there are different rules for heterogeneous or homogeneous processors. For example, sharing of load in the distribution of work in such a manner that all processors have an equal amount to do work. In this each processor can schedule from a queue that is ready common or can use an arrangement by master slave.

- ☐ SSTF is quite common and so naturally, it has a appeal
- ☐ The performance of SCAN and C-SCAN is better for system, which places a heavy load on the disk.
- ☐ Performances depend on the types & numbers of requests,

## 4. Selection of Algorithm:

To determine a particular algorithm, predetermined workload and the performance of each algorithm for that workload is to be determined. It can be said that which in turn are influenced by the file-allocation method.

- ☐ The algorithm must be written as a separate module of the operating system. It must be allowed to be replaced with other one, if necessary.
- ☐ For default, either SSTF or LOOK is a reasonable choice.

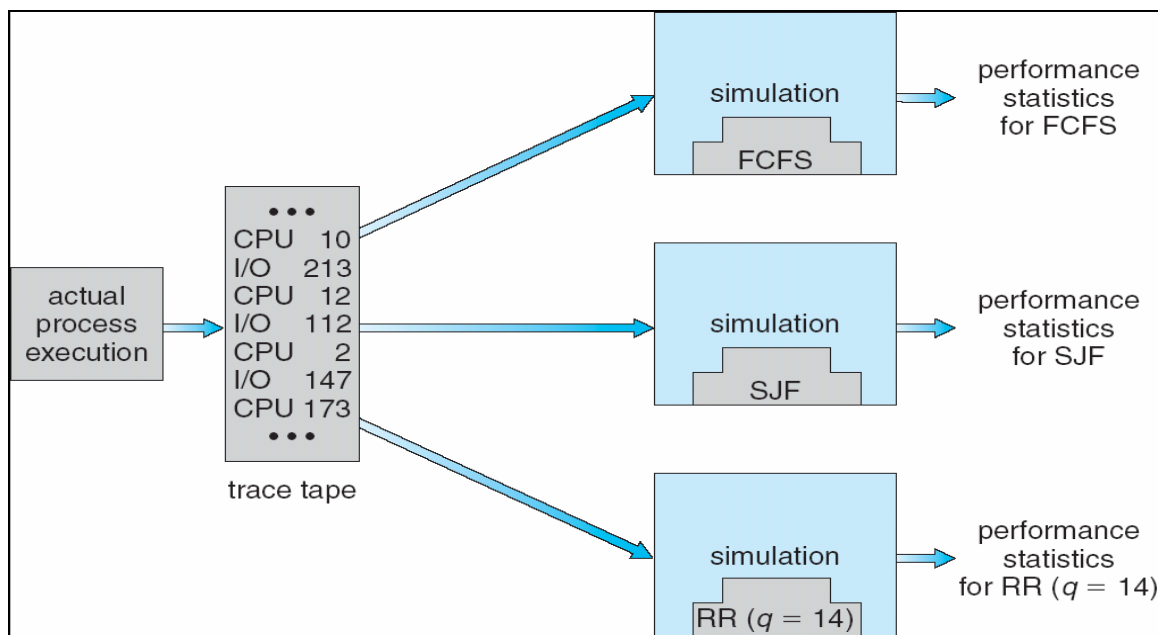

## 5. Conclusion:

A number of different scheduling algorithms have been discussed and which one is the best to work that depends on the application of it. The following table shows the comparison of different types of algorithms (starting at track 100):

| FIFO | | SSTF | | SCAN | | C-SCAN | |
|---|---|---|---|---|---|---|---|
| Next track Accessed | Number of tracks traversed | Next track accessed | Number of tracks traversed | Next track Accessed | Number of tracks traversed | Next track Accessed | Number of tracks traversed |
| 55 | 45 | 90 | 10 | 150 | 50 | 150 | 50 |
| 58 | 03 | 58 | 32 | 160 | 10 | 160 | 10 |
| 39 | 19 | 55 | 03 | 184 | 24 | 184 | 24 |
| 18 | 21 | 39 | 16 | 90 | 94 | 18 | 166 |
| 90 | 72 | 38 | 01 | 58 | 32 | 38 | 20 |
| 160 | 70 | 18 | 20 | 55 | 03 | 39 | 01 |
| 150 | 10 | 150 | 132 | 39 | 16 | 55 | 16 |
| 38 | 112 | 160 | 10 | 38 | 01 | 58 | 03 |
| 184 | 146 | 184 | 24 | 18 | 20 | 90 | 32 |
| Avge seek | 27.5 | Avge seek | 27.5 | Avge seek | 27.8 | Avge seek | 35.8 |

The OS with general purpose may use FCFS, CSCAN, preemptive and the OS with real time can opt for priority, no preemptive as in these OS performance is never obvious and Benchmarking is everything. The three types of scheduling decisions taken by OS with respect to the execution of process are:

Long term: finds when new processes are to be admitted to the system.
:

Medium term: finds when a program is bought into main memory for execution.

Short term: finds which ready process will be executed next by the processor.

The choice of algorithm depends on expected performance and on implementation complexity as shown below

| Name | Description | Remarks |
|---|---|---|
| Selection according to requester | | |
| RSS | Random scheduling | For analysis and simulation |
| FIFO | First in First Out | Fairest of them all |
| PRI | Priority by process | Control outside of queue management |
| LIFO | Last in First Out | Maximize locality and resource utilization |
| Selection according to requested ITEM | | |
| SSTF | Shortest service time first | High utilization, small queues |
| SCAN | Back and forth over disk | Better service distribution |
| CSCAN | One way with fast return | Lower service variability |
| N-step-SCAN | Scan of N records at a | Service Guarantee |

| | time | |
|---|---|---|
| FSCAN | N step Scan with N=queue size at beginning of cycle | Load sensitive |